\documentclass[english,11pt]{article}
\usepackage{amsmath,amsfonts,amssymb,amsbsy,amstext}
\usepackage[top=3cm, bottom=2cm, left=2cm, right=2cm]{geometry}
\usepackage{graphicx}
\usepackage{color}
\usepackage[english]{babel}
\usepackage{subfigure}
\usepackage[numbers,compress]{natbib}

\newcommand{\beq}{\begin{eqnarray}}
\newcommand{\eeq}{\end{eqnarray}}
\newcommand{\non}{\nonumber\\}

\newcommand{\p}{\partial}
\newcommand{\Tr}{{\rm Tr}}

\newcommand{\sign}{{\rm sign}}

\begin{document}

\begin{titlepage}
\def\thefootnote{\fnsymbol{footnote}}
\phantom{.}\vspace{-2cm}
\begin{flushright}
NORDITA-2013-96
\end{flushright}

\bigskip

\begin{center}
{\Large {\bf Baryonic sphere: 
a spherical domain wall carrying baryon number}}

\bigskip
{\large Sven Bjarke Gudnason${}^1$\footnote{\texttt{sbgu(at)kth.se}} 
and Muneto Nitta${}^2$\footnote{\texttt{nitta(at)phys-h.keio.ac.jp}}}
\end{center}

\renewcommand{\thefootnote}{\arabic{footnote}}

\begin{center}
\vspace{0em}
{\em {${}^1$Nordita, KTH Royal Institute of Technology and Stockholm University,
Roslagstullsbacken 23, SE-106 91 Stockholm, Sweden\\
${}^2$Department of Physics, and Research and Education Center for Natural
Sciences, Keio University, Hiyoshi 4-1-1, Yokohama, Kanagawa 223-8521,
Japan 

\vskip .4cm}}

\end{center}

\vspace{1.1cm}

\noindent
\begin{center} {\bf Abstract} \end{center}

We construct a spherical domain wall which has baryon charge
distributed on a sphere of finite radius in a Skyrme model with a
sixth order derivative term and a modified mass term. Its distribution
of energy density likewise takes the form of a sphere. In order to
localize the domain wall at a finite radius we need a negative
coefficient in front of the Skyrme term and a positive coefficient of
the sixth order derivative term to stabilize the soliton. Increasing
the pion mass pronounces the shell-like structure of the
configuration.

\vfill

\begin{flushleft}
{\today}
\end{flushleft}
\end{titlepage}

\hfill{}
\setcounter{footnote}{0}

\section{Introduction}

Domain walls appear in many different theories if they possess
degenerate and discrete vacua. 
They appear in field theory \cite{Manton:2004tk} 
including 
supersymmetric field theory \cite{Eto:2006pg,Shifman:2007ce}, 
quark matter \cite{Eto:2013hoa},
cosmology \cite{Vilenkin:2000}, and 
various condensed matter systems \cite{Volovik2003}.

By the scaling argument known as Derrick's theorem 
\cite{Derrick:1964ww}, a localized,
finite-energy scalar configuration with only a standard kinetic term
and a potential can only exist in $1+1$ dimensions. This can be
side-stepped by making the scalar field dependent on only 1 spatial
coordinate. If we now contemplate compactifying this configuration to
a 3-sphere (in $3+1$ dimensions), the above scaling argument tells us
that it will shrink to a point. This can be avoided by adding higher
derivative terms, like for instance the Skyrme term. The configuration
just described is simply a Skyrmion \cite{Skyrme:1961vq},
but with a modified mass term -- first introduced in the
baby Skyrme model in $d=2+1$ dimensions \cite{Kudryavtsev:1997nw} and
slightly later in the Skyrme model 
\cite{Piette:1997ce} (see also
\cite{Kopeliovich:2005vg}) -- namely a mass term for the pion fields
with two degenerate discrete vacua. 

Let us look at the configuration from the point of view of it being a
domain wall (the higher derivative terms are simply there to stabilize
its size), and describe it in terms of a four-vector $\mathbf{n}$
and potential $\tfrac{1}{2}m^2(1-n_4^2)$. 
The model admits two discrete degenerate vacua 
$n_4=-1$ and $n_4=+1$ with unbroken $SO(3)$-symmetry,
and a domain wall interpolating between them.
This domain wall possesses $S^2$ moduli 
because it breaks the vacuum symmetry down to $SO(2)$  
\cite{Losev:2000mm,Nitta:2012wi,Nitta:2012rq}.
Our configuration is simply a
configuration with (say) $n_4=-1$ at the origin and $n_4=+1$ at
spatial infinity. 
If in turn we make the $S^2$ moduli wind along 
the world volume of the domain wall, which is
also $S^2$, then remembering the radial ``winding'' along the domain
wall, it carries a topological charge $\pi_3(S^3)$, namely
a Skyrmion or baryon charge.

In fact, what we have just described is simply the traditional
Skyrmion with the addition of a modified potential for the pions. Now 
topologically speaking everything checks out, but intuitively or
physically, the so-called vacuum inside the soliton is point-like and
furthermore the energy density does not vanish at said point. In this
note we modify the model by including a higher derivative term than
the Skyrme term (i.e.~a sixth order derivative term) and show by
choosing a negative sign for the Skyrme term, keeping the coefficient
of the sixth order term positive, that the domain wall with winding
moduli can be physically pushed out from the origin and thus really
resembling a domain wall. Interestingly, the latter description is a
model proposed by Jackson et.~al.~\cite{Jackson:1985yz} while their
motivation was to make the interaction of what is interpreted as the
$\omega$-meson and using the Skyrme term to make scalar exchange
attractive (whereas in the original Skyrme model the central potential
of nucleon-nucleon interaction is all repulsive).

A hand-waving explanation of how our domain wall works
is as follows. We choose not to touch the sign of the standard
kinetic term and by Derrick's theorem, the highest derivative term
needs to have a positive coefficient. Let us further consider the
situation in which the pion mass is very large compared to other
scales in the system. 
If all the derivative terms
have positive coefficients, the cheapest way energy-wise, to
interpolate the two vacua is for the chiral angle function to make a
steep descent just at the origin. This makes the standard solutions
have the energy peak at the center.
Considering two, four and six derivative terms with a negative sign
for only the fourth order term, the cancellation between the terms
allows for the transition between the vacua to be moved to a higher
radius.

The spherical domain wall in our model 
is a $3+1$ dimensional generalization of a $2+1$ dimensional model;
an $O(3)$ nonlinear sigma model admitting two discrete degenerate 
vacua and a domain wall with a $U(1)$ modulus interpolating between
these vacua \cite{Abraham:1992vb}.\footnote{The Skyrme model admits also exact
  domain wall solutions of non-topological nature, which we do not
  consider here \cite{Canfora:2013osa}. } 
If one makes a closed domain wall with the $S^1$ modulus twisted 
along the $S^1$ world-volume, it is a lump 
\cite{Polyakov:1975yp} or baby Skyrmion \cite{Piette:1994ug}
with a topological charge of $\pi_2(S^2)$
\cite{Kobayashi:2013ju}.

In the next section we will start by reviewing the Skyrme model with a
modified mass term allowing for two discrete and degenerate vacua.

\section{The Skyrme model with a modified mass term}

Let us consider the Skyrme model \cite{Skyrme:1961vq}
\beq
\mathcal{L} = 
\frac{f_{\pi}^2}{16}\Tr\left(\p_\mu U^\dag \p^\mu U\right)
+\frac{1}{32e^2}\Tr\left([U^\dag\p_\mu U,U^\dag\p_\nu
  U]^2\right)
- V(U) \, ,
\eeq
where $f_\pi$ is the pion decay constant, $e$ is a coupling constant,
$U$ is an $SU(2)$-valued matrix field and $\mu=0,1,2,3$ runs
over 4-dimensional spacetime indices. 
Instead of the usual mass term, $\propto\Tr[2\mathbf{1}_2-U-U^\dag]$,
with only one vacuum, we consider a modified mass term which allows
for domain walls, as two vacua are present
\cite{Piette:1997ce} 
\beq
V(U) = \frac{m^2 e^2 f_\pi^4}{256}
\Tr\left[(2\mathbf{1}_2-U-U^\dag)(2\mathbf{1}_2+U+U^\dag)\right] \, ,
\eeq
where $m$ is the (here dimensionless) pion mass and $\mathbf{1}_2$ is
the two-by-two unit matrix. 
Introducing a field $n$ such that
\beq
U = i n_a \sigma^a + n_4 \mathbf{1}_2 \equiv 
\mathbf{n} \cdot \mathbf{t} \, , 
\eeq
where $a=1,2,3$ is summed over, $\sigma^a$ are the Pauli matrices and
$U^\dag U = \mathbf{1}_2$ is equivalent to
$\mathbf{n}\cdot\mathbf{n}=1$, we obtain the $O(4)$ sigma model with the
Skyrme term
\begin{align}
\mathcal{L} &= 
\frac{1}{2}\p_\mu\mathbf{n}\cdot\p^\mu\mathbf{n}
+\frac{1}{4}\left(\p_\mu\mathbf{n}\cdot\p_\nu\mathbf{n}\right)
\left(\p^\mu\mathbf{n}\cdot\p^\nu\mathbf{n}\right)
-\frac{1}{4}\left(\p_\mu\mathbf{n}\cdot\p^\mu\mathbf{n}\right)^2 
- V(\mathbf{n}) \, , \label{eq:LO4} \\
V(\mathbf{n}) &= \frac{1}{2} m^2(1-n_4^2) \, ,
\end{align}
where we have rescaled the coordinates 
$x^\mu\to\frac{2}{ef_{\pi}}x^\mu$ and the energy is given in units of
$f_{\pi}/(2e)$. 
The Skyrmion number is given by the 3rd homotopy group of the 3-sphere
and reads
\beq
B = -\frac{1}{24\pi^2}\int d^3x\; 
\epsilon_{i j k}\Tr\left(U^\dag \p_i U U^\dag \p_j U U^\dag \p_k
U\right)
= -\frac{1}{12\pi^2}\int d^3x\;
\epsilon_{i j k}\epsilon^{a b c d}
\p_i n^a \p_j n^b \p_k n^c n^d \, . 
\eeq
In order to find a spherical solution we proceed by reviewing the
standard Hedgehog Ansatz in the next section. 

\section{The Skyrmion with a modified mass term}

The most naive attempt of constructing the spherical domain wall is to
simply use the standard Hedgehog Ansatz for the Skyrmion (in the
standard way of constructing the Skyrmion solution) and study the
solutions and energy densities as functions of the mass parameter $m$,
which is the parameter controlling the width of the domain wall.  
The Hedgehog Ansatz reads
\beq
U = \exp\left\{i f(r) \hat{x}^i \sigma^i\right\} 
= \mathbf{1}_2 \cos f(r) + i \hat{x}^i \sigma^i \sin f(r) \, ,
\eeq
which in terms of $\mathbf{n}$ is
\beq
n^i = \hat{x}^i \sin f(r) \, , \qquad
n^4 = \cos f(r) \, , \label{eq:nhedgehog}
\eeq
for which the Lagrangian density reads
\beq
-\mathcal{L} =
\frac{1}{2} f_r^2 
+ \frac{1}{r^2}\sin^2 f\left(1 + f_r^2\right)
+ \frac{1}{2r^4}\sin^4 f 
+ \frac{1}{2}m^2\sin^2 f \, ,
\label{eq:LSkyrmion_sph}
\eeq
where $f_r\equiv\partial_r f$ and the energy (Skyrmion mass) is
\beq
E = 2\pi \int dr \left\{
r^2 f_r^2 
+ 2\sin^2 f\left(1 + f_r^2\right)
+ \frac{1}{r^2}\sin^4 f 
+ r^2 m^2 \sin^2 f
\right\} \, .
\label{eq:sk1en}
\eeq
The equation of motion reads
\beq
\left(r^2 + 2\sin^2 f\right) f_{rr} + 2 r f_r
+ \sin 2f \left(f_r^2 - 1 - \frac{1}{2} r^2 m^2 
- \frac{\sin^2 f}{r^2} \right) = 0 \, . \label{eq:eom_Skyrmion} 
\eeq
If we apply the boundary conditions $f(\infty)=0$ and $f(0)=\pi$, the
Skyrmion solution corresponds to a domain wall with vacuum $n_4 = 1$
at spatial infinity and vacuum $n_4 = -1$ on the inside. 
The volume of the region with vacuum $n_4 = -1$ is expected to be
point-like. 
Let us calculate the Skyrmion number of this configuration
\beq
B = - \frac{2}{\pi} \int dr \; \sin^2(f) f_r
= -\frac{1}{2\pi} \int dr\; \p_r 
\left(2f - \sin 2f\right) = \frac{f(0) - f(\infty)}{\pi} = 1 \, ,
\eeq
where the last equality depends on the boundary conditions and is true
for the above given ones. 

\begin{figure}[!tp]
\begin{center}
\mbox{\subfigure[Profile function $f$]{\includegraphics[width=0.45\linewidth]{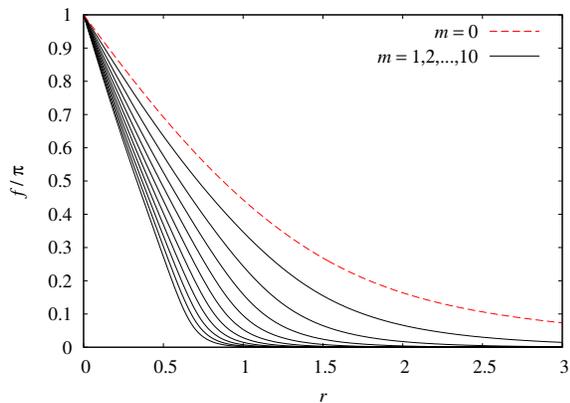}}\qquad
\subfigure[Energy density]{\includegraphics[width=0.45\linewidth]{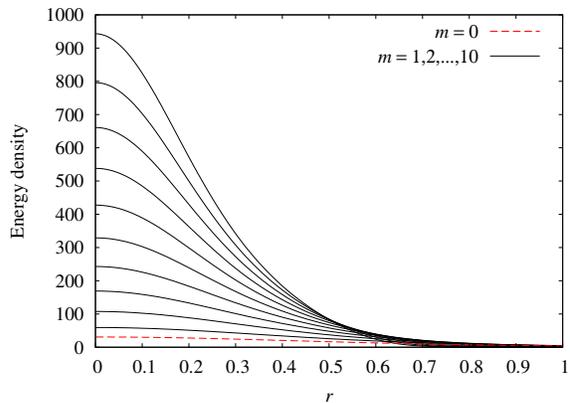}}}\\
\mbox{\subfigure[Baryon number density]{\includegraphics[width=0.45\linewidth]{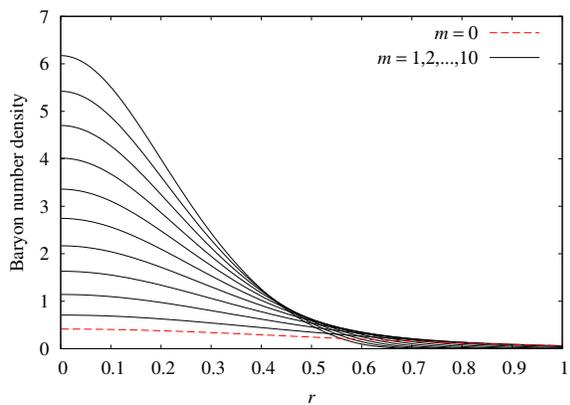}}\qquad
\subfigure[Width of domain wall]{\includegraphics[width=0.45\linewidth]{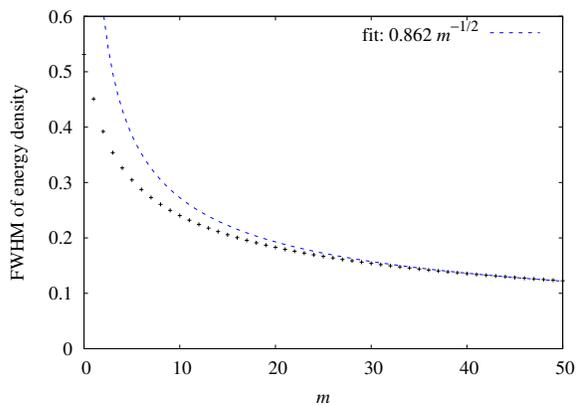}}}
\caption{(a) Profile function, (b) energy density and (c) baryon
  number density of the spherical domain wall for
  $m=0,1,2,\ldots,10$. (d) shows a full-width-half-maximum estimate of
  the width of the domain wall. The fit shows that the simple scaling
  estimate, $w\sim 3^{-1/4}/\sqrt{m}$ is a quite good approximation at
  large $m$. } 
\label{fig:sk1}
\end{center}
\end{figure}

In fig.~\ref{fig:sk1} is shown the spherical domain wall (i.e.~the
Skyrmion with the modified mass term) for $m=0,1,2,\ldots,10$. 
The energy density grows with $m$ near the center of the Skyrmion
and the size of the domain wall shrinks as $m$ increases. 
The volume of the vacuum of the core of the Skyrmion does not grow and
remains point-like for any value of the mass $m$. 
The ``vacuum'' on the inside of the domain wall is not a vacuum in the
traditional sense of the word. That is, the energy density does not
exhibit a minimum at said point.  

In fig.~\ref{fig:sk1}d is shown a numerical estimate of the width of
the domain wall. We can estimate the scaling in the limit of large
$m\gg 1$ by rescaling $r\to\mu r$ in eq.~\eqref{eq:sk1en}, neglecting
the terms proportional $1/\mu$ (for large $m$, the terms proportional
to $1/\mu$ are negligible compared to the mass term which is
proportional to $1/\mu^3$) and then varying with respect to $\mu$:
\beq
\mu = 3^{\frac{1}{4}}\sqrt{m} 
\left(\frac{\int dr\; \sin^2 f}{\int dr\left(2\sin^2(f)f_r^2 +
  \frac{1}{r^2}\sin^4 f\right)}\right)^{\frac{1}{4}} \, ,
\eeq
giving the naive scaling estimate of the domain wall width $w \sim
3^{-\frac{1}{4}}/\sqrt{m}$.

In the next section we will attempt to ``open up'' the inner vacuum
such that it is evident that we really are studying a compactified
domain wall.

\section{Higher-derivatives Skyrmion: a spherical domain wall}

In this section we consider a way of ``opening up'' the inner vacuum,
such that it is evident from the energy density that the  
deformed Skyrmion is really a spherical domain wall, namely we will
add even higher derivative terms than the Skyrme term to the
Lagrangian, keeping spherical symmetry intact.
The simplest possible extension is to add just a sixth-order
derivative term to the action. For simplicity, we will only consider
the type of higher-derivative terms that gives rise to a second-order
equation of motion, along the lines of Marleau
\cite{Marleau:1989fh}.
The action can be written as \cite{Jackson:1985yz}
\beq
\mathcal{L} = c_2 \mathcal{L}_2 + c_4 \mathcal{L}_4 + c_6
\mathcal{L}_6 - V \, ,
\eeq
with
\begin{align}
-\mathcal{L}_2 &= 
-\frac{1}{4} \Tr\left(\partial_\mu U^\dag \partial^\mu U\right)
= \frac{1}{2} f_r^2 + \frac{1}{r^2}\sin^2 f \, , \\
-\mathcal{L}_4 &= 
-\frac{1}{32}\Tr\left([U^\dag \partial_\mu U,U^\dag \partial_\nu U]^2\right)  
= \frac{1}{r^2}\sin^2(f) f_r^2
+ \frac{1}{2r^4}\sin^4 f \, , \\
-\mathcal{L}_6 &= -\frac{1}{144}
\left(\epsilon^{\mu\nu\rho\sigma}\Tr\left[U^\dag\partial_\nu U U^\dag
  \partial_\rho U U^\dag \partial_\sigma U\right]\right)^2
= \frac{1}{r^4}\sin^4(f) f_r^2 \, , \\
V &= \frac{1}{16} m^2 
\Tr\left[(2\mathbf{1}_2-U-U^\dag)(2\mathbf{1}_2+U+U^\dag)\right]
=\frac{1}{2} m^2 \sin^2 f \, .
\end{align}
Let us first consider the corresponding energy functional
\beq
E = c_2 e_2 + c_4 e_4 + c_6 e_6 + m^2 v \, , \qquad 
e_n\equiv -\int d^3x\;\mathcal{L}_n \, , \qquad 
v\equiv \frac{1}{m^2} \int d^3x\; V \, ,
\eeq
and perform a scale transformation, $x^\mu\to{x'}^{\mu}=\mu x^\mu$:
\beq
E(\mu) = \frac{c_2}{\mu} e_2 + \mu c_4 e_4 + \mu^3 c_6 e_6 +
\frac{1}{\mu^3} m^2 v \, , \label{eq:Emu}
\eeq
then according to Derrick's theorem, $E'(\mu)=0$ must have a
\emph{real} and \emph{finite} solution for $\mu$ in order for the
soliton (wall) to have a finite size (note that $\mu\to\infty$
corresponds to the soliton shrinking to a point).
A necessary but not sufficient condition for having a finite-sized
soliton solution is that the coefficient of the highest derivative
term is positive; in this case $c_6>0$. $c_4$ can have either sign.
From eq.~\eqref{eq:Emu} we can see that $e_2$ and $v$ shrink the
soliton and $e_{4,6}$ make the soliton grow if $c_4>0$. If $c_4<0$
then $e_4$ also tends to shrink the solution. 

Interestingly enough, if we choose the negative sign for $c_4$, which
means that only the sixth derivative term prevents the soliton from
shrinking to a point, the model is basically that proposed by
\cite{Jackson:1985yz} in which the authors choose the sign on
phenomenological grounds. Their motivation lies in simulating
attractive scalar exchange whereas our motivation is to study the full
parameter space of the model and find a region where the energy
density is concentrated in a shell-like structure. 

We can fix two of the coefficients by fixing the units of the length
scale and the energy scale. Let us explicitly scale the energy by
sending $E\to\lambda E$ and fixing $\mu=\sqrt{c_2/c_4}$ and
$\lambda=\sqrt{c_2|c_4|}$, which leaves us with 2 free parameters:
$c_6'=c_2c_6/c_4^2$ and $m'=m\sqrt{|c_4|}/c_2$ as well as the sign of
$c_4$ (dropping the primes)
\beq
E = e_2 + \epsilon e_4 + c_6 e_6 + m^2 v \, , \qquad
\epsilon \equiv \sign(c_4) \, .
\eeq
We thus have
\beq
-\mathcal{L} = 
\frac{1}{2} f_r^2 
+ \frac{1}{r^2}\sin^2 f 
+ \epsilon\frac{1}{r^2}\sin^2(f) f_r^2
+ \epsilon\frac{1}{2r^4}\sin^4 f
+ \frac{c_6}{r^4}\sin^4(f) f_r^2
+ \frac{1}{2} m^2 \sin^2 f \, ,
\eeq
giving rise to the equation of motion
\begin{align}
f_{rr} + \frac{2}{r} f_r + \epsilon\frac{2}{r^2}\sin^2(f) f_{rr}
-\frac{1}{r^2} \sin 2f \left[1-\epsilon f_r^2\right] - \frac{1}{2}m^2\sin 2f
+\frac{2c_6}{r^4}\sin^4 f\left[f_{rr} - \frac{2}{r} f_r\right] \non
+\frac{1}{r^4}\sin 2f\sin^2 f\left[-\epsilon1+2c_6 f_r^2\right] = 0 \, ,
\end{align}
which we will solve with the boundary conditions $f(0)=\pi$ and
$f(\infty)=0$.

In order to study the ``vacuum'' near $r=0$, let us expand the chiral
angle function $f$ as
\beq
f = \pi + f_1 r + \frac{1}{3!} f_3 r^3 + \frac{1}{5!} f_5 r^5
+ \mathcal{O}(r^7) \, ,
\eeq
where the would-be $f_{2,4,6}$ vanish due to the equation of motion. 
Plugging this expansion into the energy density yields
\begin{align}
\mathcal{E} &= \frac{1}{2} f_1^2\left[3 
+ 3\epsilon f_1^2 + 2c_6 f_1^4\right] 
+ \frac{1}{6} f_1 \left[
  3 m^2 f_1 
  - 2f_1^3 
  - 4\epsilon f_1^5 
  - 4c_6 f_1^7 
  + 5f_3
  + 10\epsilon f_1^2 f_3
  + 10c_6 f_1^4 f_3 \right] r^2 
\non &\phantom{=\ }
+ \frac{1}{360} \big[
  - 60m^2 f_1^4 
  + 16f_1^6 
  + 52\epsilon f_1^8  
  + 72c_6 f_1^{10} 
  + 60m^2 f_1 f_3
  - 80f_1^3 f_3 
  - 320\epsilon f_1^5 f_3  
  - 480c_6 f_1^7 f_3 
\non &\phantom{=+ \frac{1}{360} \big[\ }
  + 250\epsilon f_1^2 f_3^2 
  + 55f_1^3 f_3^2
  + 390 c_6 f_1^4 f_3^2 
  + 21 f_1 f_5
  + 42\epsilon f_1^3 f_5 
  + 42c_6 f_1^5 f_5
  \big] r^4
  + \mathcal{O}(r^6) \, , \label{eq:energydensity_expansion}
\end{align}
which means that a sufficient condition for the energy density to
vanish around $r\to 0$ is that the first derivative vanishes at
$r=0$ (see the equations of motion below). Using the equation of
motion at order $\mathcal{O}(r)$ and $\mathcal{O}(r^3)$, we can
determine the third and fifth derivative in terms of $f_1$
\begin{align}
f_3 &= \frac{4c_6 f_1^7 - 2\epsilon f_1^5 - 4f_1^3 + 3f_1
  m^2}{5\left(2c_6 f_1^4 + 2\epsilon f_1^2 + 1\right)} \, , \\
f_5 &= \frac{
  - 20m^2 f_1^3
  + 8f_1^5
  + 16\epsilon f_1^7 
  - 24c_6 f_1^9
  + 5f_3 m^2
  - 20f_1^2 f_3
  + 30\epsilon f_1^4 f_3
  + 140c_6 f_1^6 f_3
  - 40\epsilon f_1 f_3^2 
  - 80c_6 f_1^3 f_3^2 }
  {7\left(1 + 2\epsilon f_1^2 + 2 c_6 f_1^4\right)},
\end{align}
which we can insert into the energy density
\eqref{eq:energydensity_expansion}. Since we are interested in the
expression for small $f_1\ll 1$, we give the expression to just fourth 
order in $f_1$:
\beq
\mathcal{E} = 
\frac{3}{2} \left(f_1^2 + \epsilon f_1^4\right)
+ \left(m^2 f_1^2 - f_1^4\right) r^2 
+ \left[\frac{9}{50} m^4 f_1^2 
- \left(\frac{22}{25} m^2 + \frac{17}{50}\epsilon m^4\right) f_1^4
  \right]r^4 
+ \mathcal{O}(r^6,f_1^6) \, .
\eeq
Unfortunately, we cannot find an analytic expression for $f_1$ as it
is not a perturbative object, but encodes information about the
soliton as a whole. 
\begin{figure}[!tp]
\begin{center}
\mbox{\subfigure[$\epsilon=+$]{\includegraphics[width=0.47\linewidth]{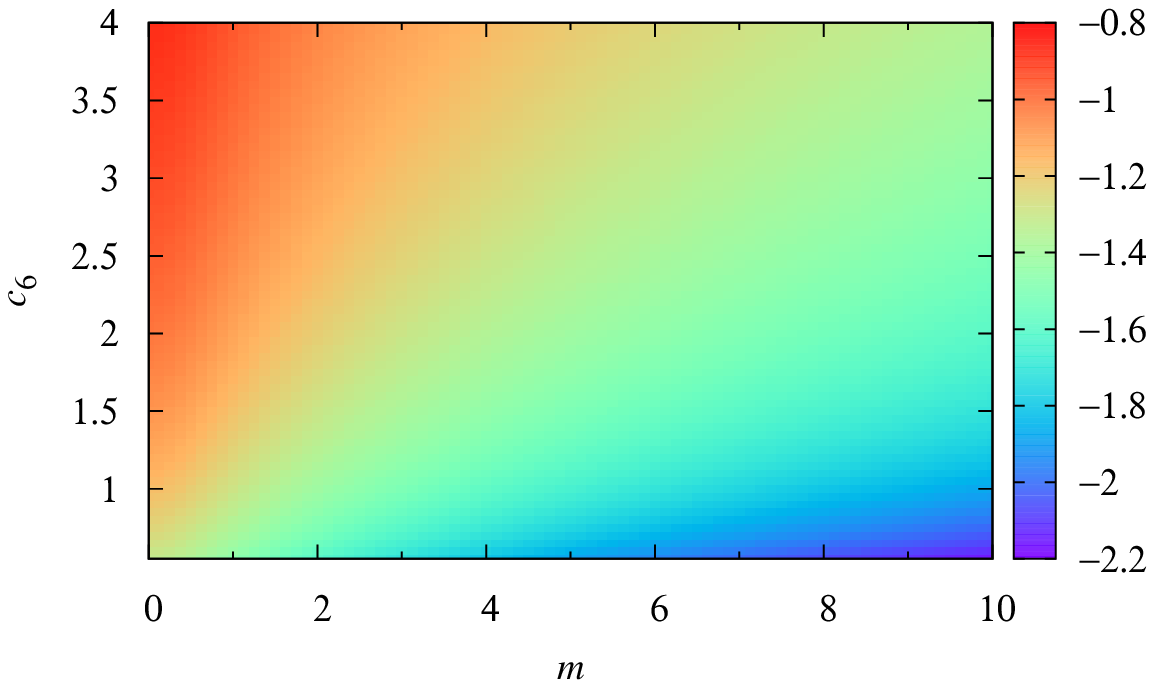}}\qquad
\subfigure[$\epsilon=-$]{\includegraphics[width=0.47\linewidth]{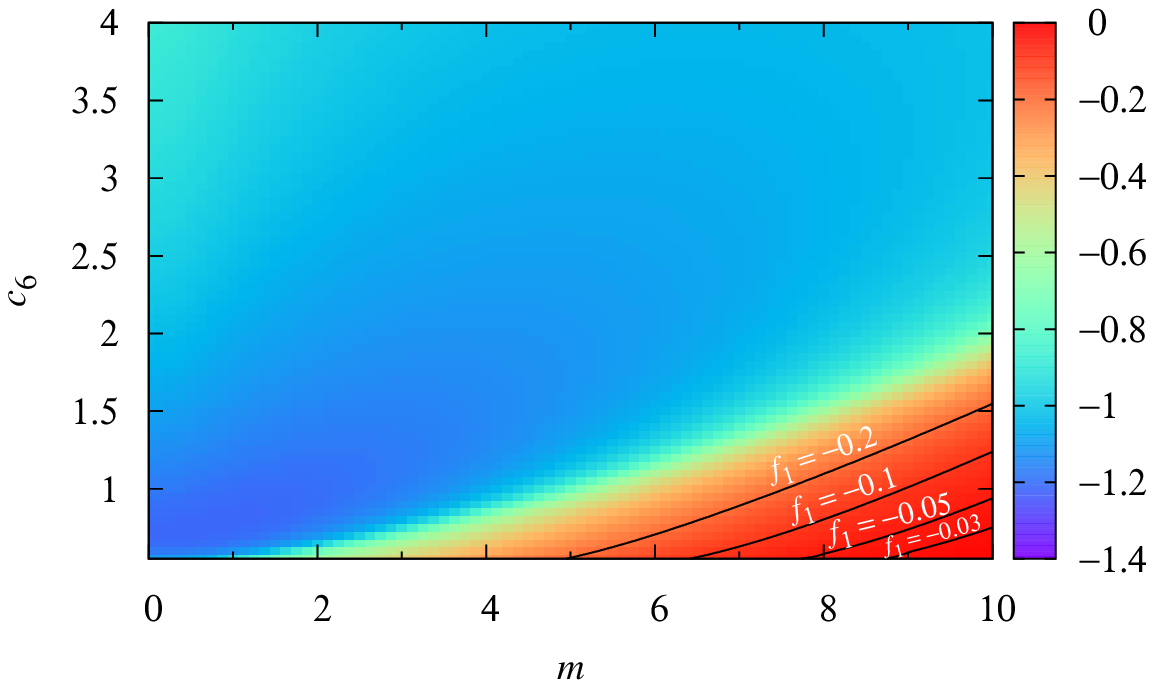}}}
\caption{The first derivative of the chiral angle function $f_1$ at
  $r=0$ as function of the pion mass $m$ and the sixth order
  derivative term coefficient $c_6$ for (a) $\epsilon=+$ and (b)
  $\epsilon=-$. In (b) four black solid lines are shown, representing
  the iso-curves $f_1=-0.2$, $-0.1$, $-0.05$ and $-0.03$ from above. }
\label{fig:f1}
\end{center}
\end{figure}
We can, however, solve the equation of motion numerically and study
$f_1$ as function of $m$ and $c_6$ for $\epsilon=\pm$ which we show in
fig.~\ref{fig:f1}.

\begin{figure}[!htp]
\begin{center}
\mbox{\subfigure[$c_6=1$]{
$\begin{array}{c}
\includegraphics[width=0.45\linewidth]{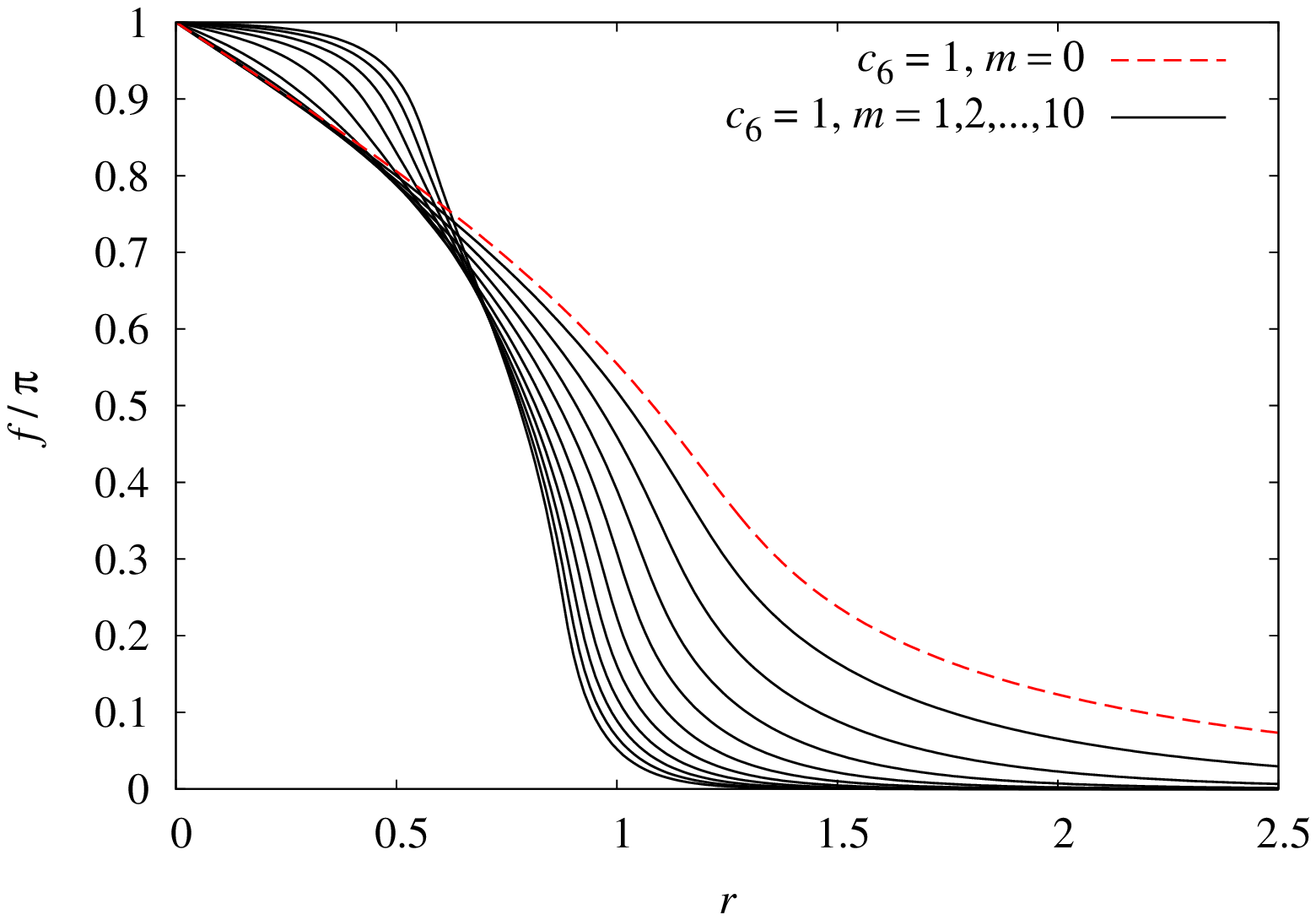}\\
\includegraphics[width=0.45\linewidth]{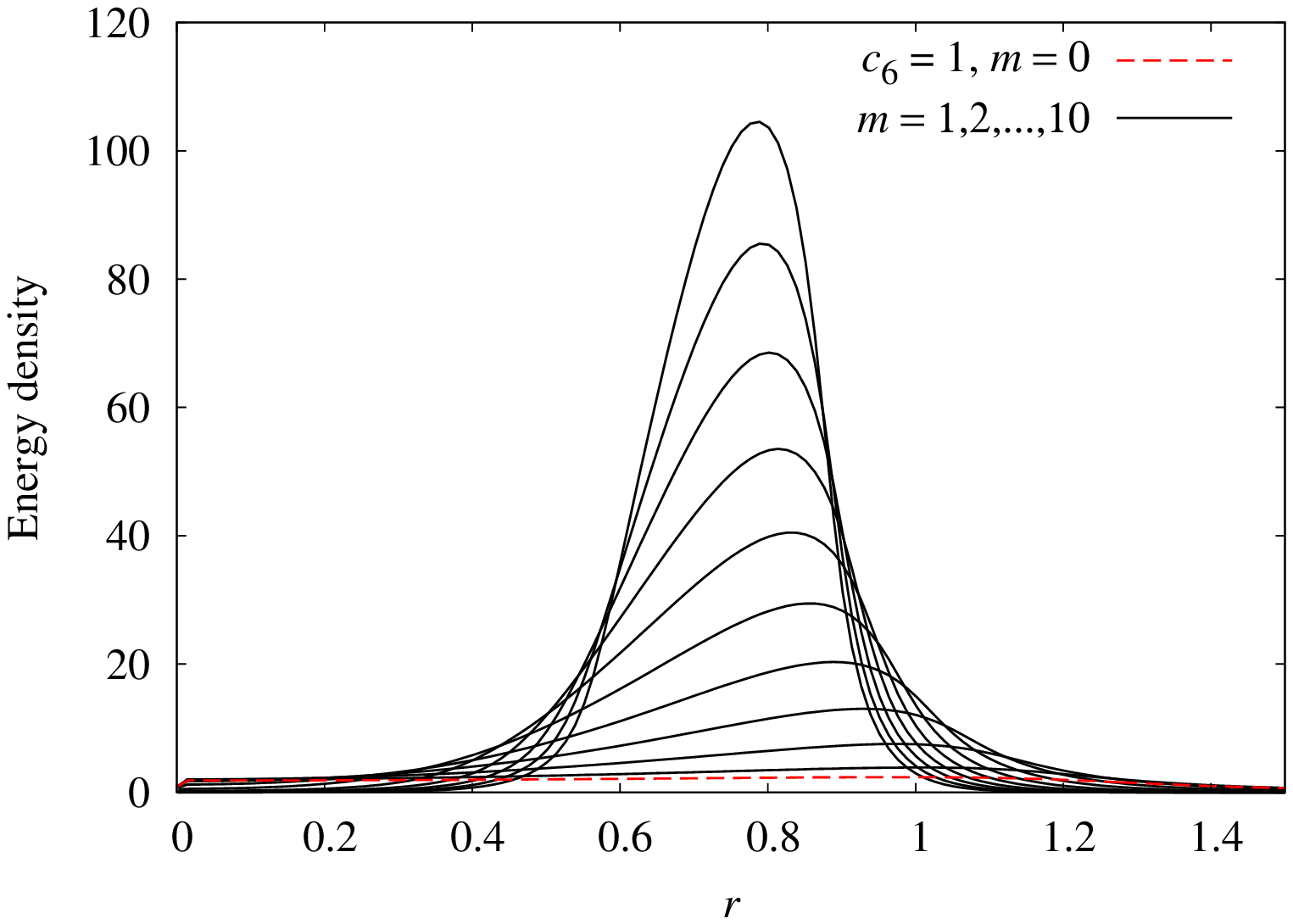}\\
\includegraphics[width=0.45\linewidth]{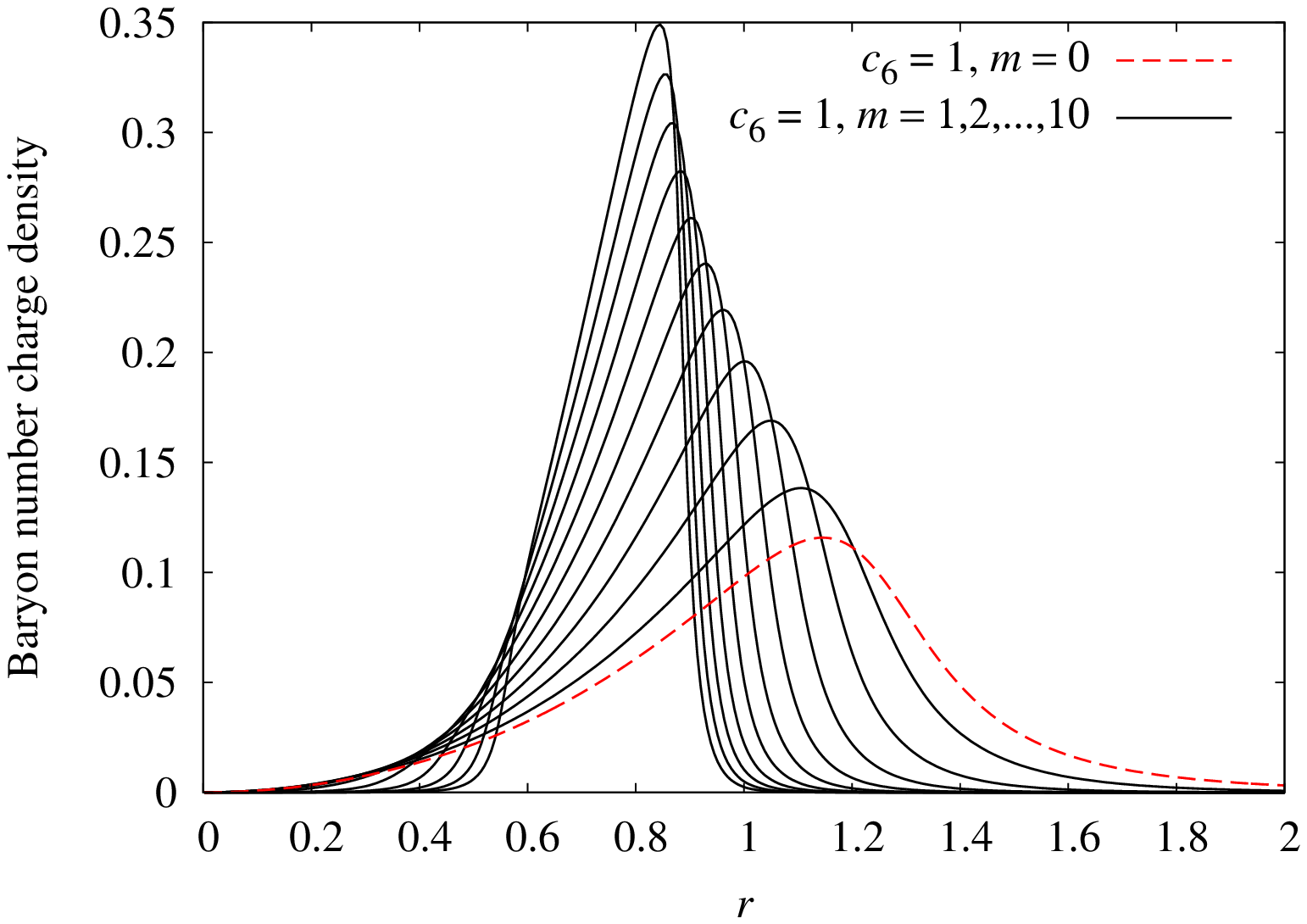}
\end{array}$}
\subfigure[$c_6=0.6$]{
$\begin{array}{c}
\includegraphics[width=0.45\linewidth]{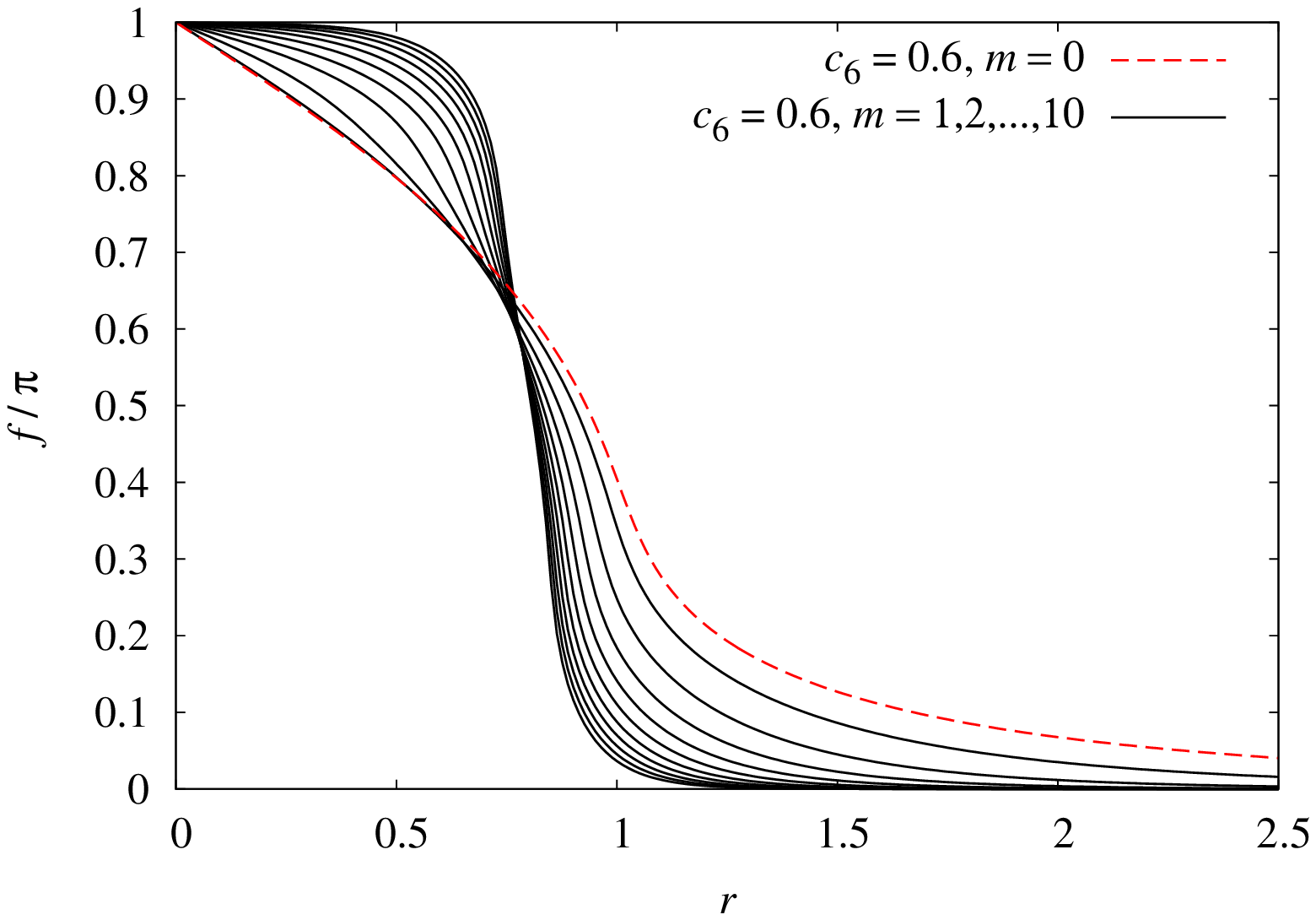}\\
\includegraphics[width=0.45\linewidth]{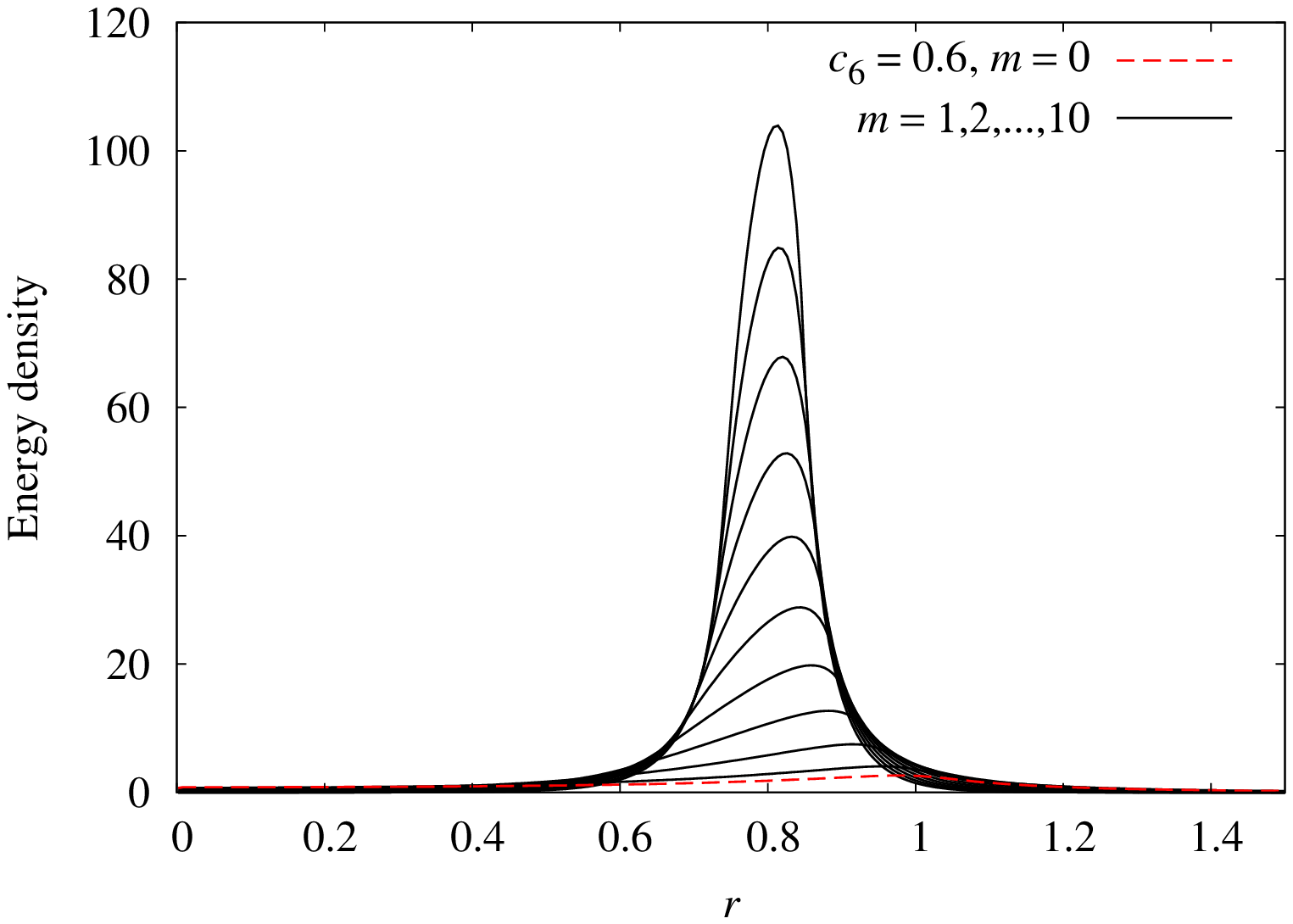}\\
\includegraphics[width=0.45\linewidth]{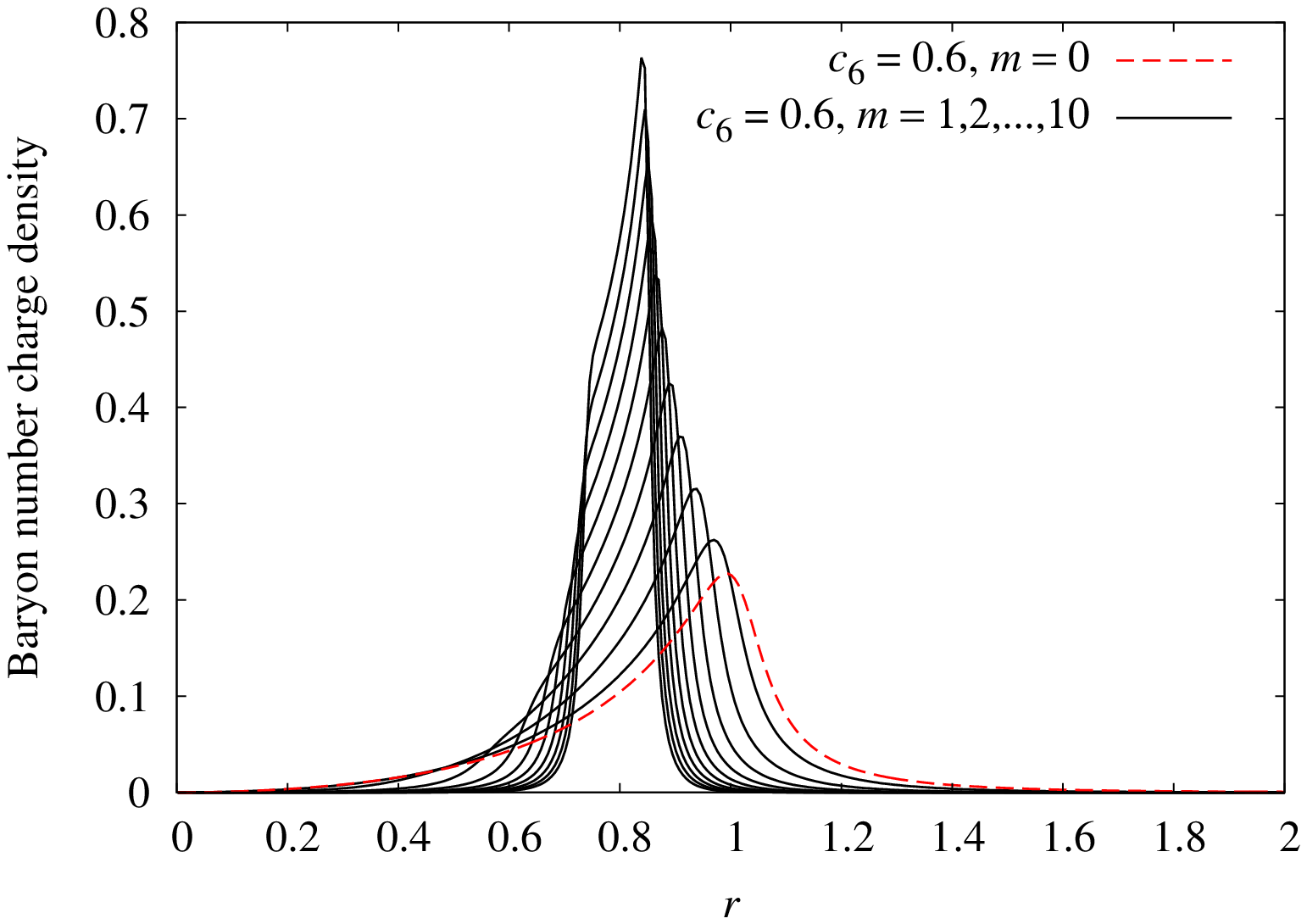}
\end{array}$}}
\caption{Sixth order derivative Skyrmions with negative Skyrme term for
$c_6=1$ (left) and $c_6=0.6$ (right) and various pion masses
  $m=0,1,\ldots 10$. }
\label{fig:hdsk}
\end{center}
\end{figure}

Finally, we can show numerical solutions in the interesting region of
parameter space, i.e.~for small $c_6$ and relatively large pion mass
$m$. 
In fig.~\ref{fig:hdsk} is shown the profile function $f$, the energy
density and the baryon number charge density for two values of
$c_6=1,0.6$ and various pion masses $m=0,1,\ldots,10$. 

\section{Summary and Discussion}

We have constructed a spherical domain wall 
with baryon charge distributed on the surface of a sphere with a finite
radius in a Skyrme model with the addition of 
a modified mass term and a sixth-order derivative term.
The width of the domain wall is inversely proportional to (the square
root of) the pion mass $m$, in units of the coefficient of the kinetic 
term. 
By means of a series expansion near the origin, we have related the
energy density at small radii to the first derivative of the chiral
angle function at the origin, $f_1$. When $f_1$ is parametrically
small, the energy as well as the baryon charge density remain
parametrically small near the origin of the soliton. This separation
of the domain wall from the origin point reveals that the lump of
energy of a Skyrmion in fact is a domain wall in disguise. 
In order to practically achieve an almost vanishing $f_1$ (the
first derivative of $f$) it is necessary to flip the sign of the
Skyrme term with respect to the conventional choice. This in turn
necessitates a positive sixth order term.

\bigskip

It is possible to obtain a similar behavior by keeping a positive
coefficient, $c_4$, of the Skyrme term by having a negative $c_6$
which in turn necessitates a positive (and sufficiently large) $c_8$,
for an eighth-order derivative term,
e.g.~\cite{Marleau:1989fh} 
\beq
-\mathcal{L}_8 = \frac{1}{r^6}\sin^6(f) f_r^2 
- \frac{1}{4r^8}\sin^8 f \, .
\eeq
In this case the balance of forces is three versus two; namely the
kinetic term, the sixth order term and the potential tend to shrink
the soliton whereas the Skyrme term and the eighth order term tend to
make the soliton grow. 

In the Skyrme model, there is no primary reason for using 
the conventional mass term (i.e.~$\tfrac{1}{2}m^2(1-n_4)$) instead of 
the modified mass term for the pion mass. 
One can also consider higher derivative terms like a sixth order term
-- as in our case -- for a model of baryons. 
Therefore, there is a possibility that baryons are spherical. 
What impact it has in nuclear physics remains a future problem. 

Interestingly, a BPS proposal which has been put forward in
\cite{Adam:2010fg}, relies on the sixth-order derivative term (only)
for saturation of the BPS bound and the near BPS region (which has
only small contributions from the second and fourth order terms) is
phenomenologically compelling because it gives a low binding energy 
and an almost linear relation between the baryon number and mass
\cite{Adam:2013tda}.

It is an open problem to construct higher winding Skyrmions. 
In particular, it is to be studied 
whether the energy distributions of the minimum-energy
configurations  
can be spherical or need to be in separated lumps.
For this purpose, the rational map Ansatz \cite{Houghton:1997kg} 
may be useful as for usual Skyrmions.\footnote{For analytic
  properties of multi-Skyrmions in the standard Skyrme model, see
  \cite{Kopeliovich:2001cg}. } 
Likewise, the interaction between spherical domain walls as 
Skyrmions is an important subject.
Another interesting topic is a coupling to gravity, 
resulting in a gravitational Skyrmion or a black hole.

It is interesting to study low-energy modes of 
a spherical domain wall. 
For a flat domain wall, the effective theory 
is a nonlinear sigma model with the target space 
${\mathbb R}\times S^2$, 
describing the fluctuations of the domain wall surface 
and the $S^2$ Nambu-Goldstone modes 
\cite{Nitta:2012wi,Nitta:2012rq}.
The effective field theory of a spherical domain wall 
may contain a radial fluctuation field which should be light 
but not massless, 
and the $S^2$ Nambu-Goldstone modes 
which are twisted. 
Low-energy effective field theories on 
curved soliton world-volumes have not been studied in depth thus far. 
This model provides a primary example of such.

When we deform the model with an additional mass term 
$V_2 = - m_3^2 n_3^2$, where $m_3 \ll m$, 
there appears a domain line inside a domain wall
\cite{Nitta:2012rq}. 
In our case, a circular domain line will appear 
in a spherical domain wall. 
If we further deform the model by adding 
$V_3 = - m_2^2 n_2$ with $m_2 \ll m_3 \ll m$, 
sine-Gordon kinks appear on the domain line 
\cite{Nitta:2012xq}. 
In our setting, sine-Gordon kinks will appear 
on a circular domain line in a spherical domain wall. 
We may construct a domain wall junction on a sphere 
\cite{Brito:2001ga}
by properly choosing the potential.

In the introduction, we mentioned that 
our spherical domain wall is a $d=3+1$ dimensional generalization of 
a circular domain wall being a baby Skyrmion 
with a topological charge of $\pi_2(S^2)$ 
in $d=2+1$ dimensions 
\cite{Kobayashi:2013ju}. 
Higher dimensional versions are also possible 
in an $O(N+1)$ sigma model in $d=N+1$ dimensions admitting two
discrete degenerate vacua  
and a domain wall with $S^{N-1}$ moduli
interpolating between these vacua \cite{Nitta:2012rq}. 
This model will admit an $S^{N-1}$ domain wall 
with a topological charge $\pi_{N} (S^{N})\simeq \mathbb{Z}$ 
in $d=N+1$ dimensions.

There is another higher-dimensional generalization.
The $O(3)$ sigma model mentioned above 
also admits domain walls in a torus shape $T^2 =S^1 \times S^1$ in
$d=3+1$ dimensions, 
along whose two cycles the $U(1)$ modulus is twisted 
\cite{Bolognesi:2007zz}. 
This toroidal domain wall carries Hopf charge characterized by
$\pi_3(S^2) \simeq \mathbb{Z}$. 
Therefore, our domain wall with an $S^2 \times S^2$ world-volume 
is a possibility in $d=5+1$ dimensions
and it may carry a topological charge of 
$\pi_5(S^3) \simeq \mathbb{Z}_2$.

In this paper, 
we have studied a spherical domain wall as a 
twisted soliton, i.e, the $S^2$ moduli are twisted along 
a closed domain wall world-volume,  
in the nonlinear sigma model. 
On the other hand, 
Yang-Mills Higgs theories with certain matter contents 
have been proposed to admit an $S^3$ domain wall, 
$S^2$ vortex sheet, or $S^1$ monopole string as 
a Yang-Mills instanton-particle
in $d=4+1$ dimensions, if the $S^3$, $S^2$ or $S^1$ moduli are twisted along the world-volume \cite{Nitta:2013vaa}. 
Thus far, we know of several examples of twisted solitons regarded as 
Skyrmions or instantons. 
There should exist a general framework for studying
which topological charges are carried by twisted solitons.

\section*{Acknowledgments}

The work of MN is supported in part by Grant-in-Aid for Scientific Research (No. 25400268) 
and by the ``Topological Quantum Phenomena'' 
Grant-in-Aid for Scientific Research 
on Innovative Areas (No. 25103720)  
from the Ministry of Education, Culture, Sports, Science and Technology 
(MEXT) of Japan. 
SBG thanks Keio University for hospitality during which this project
took shape.

\end{document}